\documentclass[prl,twocolumn,floatfix]{revtex4}

\usepackage{amsmath,amsthm,amsfonts,amssymb,bm}

\newcommand{\refsec}[1]{\S \ref{sec:#1}}

\newcommand{\wc}{{\rm wc}}

\newcommand{\btab}{\begin{tabular}}
\newcommand{\etab}{\end{tabular}}

\newcommand{\nth}{{n_\th}}

\newcommand{\crao}{Cram\'{e}r-Rao\ }

\newcommand{\nout}{{n_{\rm out}}}

\newcommand{\lamh}{\lam^{\rm opt}}

\newcommand{\Th}{\Theta}

\newcommand{\nex}{{\ell_{\expt}}}

\renewcommand{\hat}[1]{\widehat{#1}}

\newcommand{\mbf}[1]{\mbox{\boldmath $#1$}}

\newcommand{\ket}[1]{\mbf{|}#1\mbf{\rangle}}
\newcommand{\bra}[1]{\mbf{\langle}#1\mbf{|}}

\newcommand{\avg}{ {\bf E} }
\newcommand{\trace}{{\rm Tr}}

\renewcommand{\hbar}{ {h {\!\!\!^{\scriptscriptstyle -} } } }

\newcommand{\gam}{\gamma}
\newcommand{\alf}{\alpha}

\newcommand{\lam}{\lambda}
\newcommand{\bet}{\beta}
\renewcommand{\th}{\theta}

\newcommand{\sig}{\sigma}

\newcommand{\thh}{{\hat{\theta}}}

\newcommand{\eg}{\emph{e.g.}}
\newcommand{\ie}{\emph{i.e.}}

\newcommand{\bquem}{\begin{quote}\begin{em}}
\newcommand{\equem}{\end{em}\end{quote}}

\newcommand{\blist}{\begin{description}}
\newcommand{\elist}{\end{description}}

\newcommand{\bquote}{\begin{quote}}
\newcommand{\equote}{\end{quote}}

\newcommand{\ben}{\begin{enumerate}}
\newcommand{\een}{\end{enumerate}}

\newcommand{\bit}{\begin{itemize}}
\newcommand{\eit}{\end{itemize}}

\newcommand{\bea}{\begin{array}}
\newcommand{\eea}{\end{array}}

\newcommand{\bds}{\begin{displaystyle}}
\newcommand{\eds}{\end{displaystyle}}

\newcommand{\Rbf}{{\mathbf R}}

\newcommand{\ds}{\displaystyle}

\newcommand{\refeq}[1]{(\ref{eq:#1})}

\newcommand{\set}[2]{ \left\{ \,#1\, \left| \,#2\, \right.\right\} }

\newcommand{\seq}[1]{ \left\{ #1 \right\} }

\newcommand{\opt}{{\rm opt}}

\newcommand{\mathbox}[1]{
\fbox{$\ds #1 $}
}

\makeatletter
\def\beq{\@ifnextchar 
[{\@tempswatrue\@beq}{\@tempswafalse\@beq[]}}
\def\@beq[#1]{\begin{equation}\edef\@tmparg{#1}\ifx\@tmparg\@e
mpty \else
	\label{#1}\fi}
\newcommand{\eeq}{\end{equation}}
\makeatother 

\newcommand{\beqaa}{\begin{eqnarray*}}
\newcommand{\eeqaa}{\end{eqnarray*}}

\newcommand{\beqa}{\begin{eqnarray}}
\newcommand{\eeqa}{\end{eqnarray}}

\newcommand{\bc}{\begin{center}}
\newcommand{\ec}{\end{center}}

\usepackage{epsfig}
\usepackage{times}
\usepackage{psfrag}
\usepackage[dvips]{color}
\newcommand{\red}[1]{\textcolor{red}{#1}}

\newcommand{\ac}{{\rm ac}}
\newcommand{\ik}{ {i|k} }

\newcommand{\ts}{\textstyle}
\newcommand{\sumts}{\ts\sum}

\newcommand{\lamwc}{\lam^\wc}
\newcommand{\lamac}{\lam^\ac}

\newcommand{\gav}{g_{\rm avg}}

\newcommand{\un}{{\bf 1}}

\renewcommand{\lamh}{\hat{\lam}}
\newcommand{\nconfig}{{N_{\rm config}}}

\newcommand{\dimphi}{{n_\phi}}

\renewcommand{\nth}{{N_\th}}

\newcommand{\npovm}{{N_{\rm povm}}}

\newcommand{\ninput}{{N_{\rm input}}}

\newcommand{\iout}{{i}}

\renewcommand{\nout}{{N_{\rm out}}}
\renewcommand{\nex}{N}

\begin{document}

\title{
Quantum Metrology Subject to Instrumentation Constraints
}

\author{Robert L. Kosut}
\affiliation{SC Solutions, Inc., 1261 Oakmead Parkway, Sunnyvale, CA 94085}

\begin{abstract}

Maximizing the precision in estimating parameters in a quantum system
subject to instrumentation constraints is cast as a convex
optimization problem.  We account for prior knowledge about the
parameter range by developing a worst-case and average case objective
for optimizing the precision. Focusing on the single parameter case,
we show that the optimization problems are {\em linear programs}. For
the average case the solution to the linear program can be expressed
analytically and involves a simple search: finding the largest element
in a list. An example is presented which compares what is possible under
constraints against the ideal with no constraints, the Quantum Fisher
Information.

\end{abstract}

\maketitle

\section{Introduction}

The theoretical limit on the accuracy of parameter estimation in
quantum metrology applications has been examined in depth, \eg,
\cite{Holevo:82,BraunsteinC:94,SarovarM:06,GLM:06,BFCG:07,ShajiC:07}.
These studies reveal that special preparation of the instrumentation
-- the probe -- can achieve an asymptotic variance smaller than the
\crao lower bound \cite{Cramer:46}, often referred to as the
\emph{Quantum Fisher Information}, abbreviate here as QFI.  In
addition, the unique quantum property of {entanglement} can increase
the parameter estimation convergence rate for $N$ identical,
independent experiments from the shot-noise limit of $1/\sqrt{N}$ to
the Heisenberg limit of $1/N$.

It is reasonable to expect, with or without entanglement, that the QFI
will not be obtained with imperfect and limited instrumentation
resources, \ie, not all states can be prepared and not all measurement
schemes are possible. Under these conditions what exactly is the best
that can be done?

In this paper we present an approach which maximizes the parameter
estimation accuracy in the presence of limits on instrumentation,
The method is based on the convex optimization approach to optimal
experiment design as developed in \cite{BoydV:04} and as applied to
quantum tomography in \cite{KosutWR:04}. Incorporating prior knowledge
of the parameter range, we develop a worst-case and average case
objective for optimizing the precision. Focusing on the single
parameter case, we show that the optimization problems are {\em linear
programs}. For the average case the solution to the linear program can
be expressed analytically and involves a simple search, \ie, finding
the largest element in a list. This means that an enormous number of
combinations of state and sensor configurations can be efficiently
evaluated.

\section{Optimal Experiment Design}
\label{sec:oed}

Consider a quantum system dependent on an {\em unknown} scalar real
parameter $\th$ which is known \emph{a priori} to be in a set
$\Th=\set{\th}{\th_{\min}\leq\th\leq\th_{\max}}$. The parameter $\th$
is to be estimated using data from repeated {\em independent,
identical} experiments. In each experiment the system can be put in
any one of $k=1,\ldots,\nconfig$ \emph{configurations}.  These are the
available settings of input states and measurements.
Each experiment in configuration $k$ results in one of $\nout$ outcomes
with probability
$
p_\ik(\th),\ i=1,\ldots,\nout
$.
Let $N_\ik(\th)$ denote the number of times outcome $i$ is obtained
from $\nex_k$ identical experiments in configuration $k$. Thus,
$
\avg{N_\ik(\th)} = \nex_k p_\ik(\th),\
\sum_{i=1}^\nout\ N_\ik(\th) = \nex_k
$
where $\avg$ is the expected value operator with respect to the
probability distribution $p_\ik(\th)$.  Let $\nex$ denote the
total number of experiments and $\lam_k$ the {\em distribution of
experiments} in configuration $k$.  Thus,
$
\lam_k = \nex_k/\nex\
\Rightarrow\
\sum_{k=1}^\nconfig\ \lam_k = \un^T\lam = 1
$
. The problem is to select the distribution of experiments per
configuration, $\lam_k,k=1,\ldots,\nconfig$, or equivalently the
number of experiments per configuration, $\nex_k$, so as to obtain
an estimate of $\th\in\Th$ with the best accuracy from $\nex$
experiments. The ``best'' attainable estimation accuracy is defined
here as the smallest possible \crao bound on the estimation variance
\cite{Cramer:46}.

Specifically, if $\thh_\nex$ is an unbiased estimate of $\th$ from
$\nex$ data, then the estimation error variance satisfies,
\beq[eq:var th]
\bea{l}
\nex F(\lam,\th)\
\avg{(\thh_N-\th)^2}
\geq
1
\\
F(\lam,\th) = \lam^T g(\th)
=
\sum_{k=1}^\nconfig\ \lam_k g_k(\th)
\\
g_k(\th) 
=
\sum_{\iout=1}^\nout\
\Big(\nabla_\th\ p_\ik(\th)\Big)^2/p_\ik(\th)
\eea
\eeq
To achieve the best accuracy we will select $\lam$ so as to maximize a
measure of the size of the {\em Fisher Information}, $F(\lam,\th)$. To
account for the knowledge that $\th\in\Th$ we will consider two
experiment design objectives for selecting $\lam$: \emph{average case} and
\emph{worst-case}.

%
\vspace{-4ex}
\beq[eq:ac]
\bea{c}
\mbox{\bf Average-Case Experiment Design}
\\
\bea{ll}
\mbox{maximize}
&
F_\ac(\lam) = 
\lam^T\gav
\\
\mbox{subject to}
&
\un^T\lam=1,\
\mbox{$\nex\lam$ is a vector of integers}
\eea
\eea
\eeq
with $\gav=\int p(\th)g(\th)d\th$ where $p(\th)$ is the probability
density associated with $\th\in\Th$.  Although the objective function
(average Fisher information) is linear in $\lam$, the integer
constraint on $\lam$ makes the optimization problem hard. Utilizing
the optimal experiment design method presented in \cite[\S
7.5]{BoydV:04}, the integer constraint is \emph{relaxed} to the linear
inequality $\lam\geq 0$. In addition, suppose we take a finite number
of samples from the set $\Th$, say,
$
\set{\th_r}{r=1,\ldots,\nth}
$
Then the non-convex integer optimization \refeq{ac} is approximated by,
\beq[eq:aca]
\bea{ll}
\mbox{maximize}
&
F_\ac(\lam)=\lam^T\gav,
\;
\gav=\sum_r\ p(\th_r) g(\th_r)
\\
\mbox{subject to}
&
\un^T\lam=1,\
\lam\geq 0
\eea
\eeq
This is a convex optimization problem in $\lam$, in fact, it is a
\emph{linear program} (LP). However, a particular advantage of this
formulation \refeq{aca}, is that the solution is given explicitly by,
\beq[eq:lam acs1]
\bea{c}
\lamh_{k}
=
\left\{
\bea{ll}
1 
& 
k = \arg\max_{k'}\  
\sum_r\ p(\th_r) g_{k'}(\th_r)
\\
0 & \mbox{otherwise}
\eea
\right.
\eea
\eeq
with the optimal objective $F_\ac(\lamh)=\max_k\ \sum_r\ p(\th_r)
g_{k}(\th_r)$.  It is possible that there is more than one optimal
distribution because $\max_{k}$ may not be unique. However, due to
limits on numerical precision, it is more likely that there are other
choices which give similar results to the optimal objective.

%
\vspace{-4ex}
\beq[eq:wc]
\bea{c}
\mbox{\bf Worst-Case Experiment Design}
\\
\bea{ll}
\mbox{maximize}
&
F_\wc(\lam) = \min_{\th\in\Th}\ 
\lam^T g(\th)
\\
\mbox{subject to}
&
\un^T\lam=1,\
\mbox{$\nex\lam$ is a vector of integers}
\eea
\eea
\eeq
As in the average-case, relaxing the integer constraint and
approximating the objective function over a set of $\th$ sampled from
the known set $\Th$ gives the optimization problem:
\beq[eq:wca]
\bea{ll}
\mbox{maximize}
&
F_\wc(\lam) = \min_r\
\lam^T g(\th_r)
\\
\mbox{subject to}
&
\un^T\lam=1,\ \lam \geq 0
\eea
\eeq
This is also an LP in $\lam$, but unlike the average-case, there is no
explicit solution. However, it can be solved efficiently for a very
large number of configurations $\nconfig$. A potential advantage of
the average-case solution over the worst-case solution is that only a
{\em single} configuration is required. As we will see in the example
to follow, the two distributions can be quite different even though
the Fisher information is similar.

The solution to both of the relaxed and approximated problems
\refeq{aca},\refeq{wca} provide upper and lower bounds to the unknown
solution of each with the integer constraint active.  Specifically,
let $\lam^\opt$ denote a solution to either \refeq{aca} or \refeq{wca}
with the integer constraint.  Let $\lamh$ be a solution to the relaxed
(LP) versions. From the latter we can determine a nearby solution
which satisfies the integer constraint, \eg, set $\lam^{\rm rnd}={\bf
round}(\lamh)$.  Then,
$
F_\ac(\lamh)
\leq
F_\ac(\lam^\opt)
\leq
F_\ac(\lam^{\rm rnd})
$ 
and $\nex_k=\nex\lam_k^{\rm rnd}$ is the number of experiments to
repeat in configuration $k$. 

\section{Quantum system parameter estimation} 
\label{sec:qparest}

\begin{figure}[t]
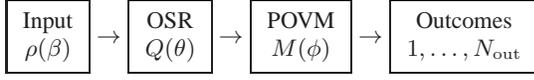

\btab{c}
$
\mathbox{
\bea{c}
\mbox{Input}
\\
\rho(\bet) 
\eea
}
\rightarrow
\mathbox{
\bea{c}
\mbox{OSR}
\\
Q(\th)
\eea
}
\rightarrow 
\mathbox{
\bea{c}
\mbox{POVM}
\\
M(\phi)
\eea
}
\rightarrow
\mathbox{
\bea{c}
\mbox{Outcomes}
\\
\mbox{\small $1,\ldots,\nout$}
\eea
}
$
\\
\\
\etab
\caption{Quantum system for estimating parameter $\th$.}
\label{fig:qsys}
\end{figure}

For the quantum system depicted in Figure \ref{fig:qsys}, the quantum
channel, $Q(\th)$, depends on the parameter $\th\in\Th$, the input
state, $\rho(\bet)$, is dependent on the input configuration parameter
$\bet$, and the POVM elements, $M_i(\phi),\ i=1,\ldots,\nout$ with
$\sum_i M_i(\phi)=I$, depend on the configuration parameter $\phi$.
Suppose that $Q(\th)$ can be described in terms of the Kraus Operator
Sum Representation (OSR) with elements $Q_k(\th)$. Then the outcome
probabilities are:
\beq[eq:pout]
\bea{rcl}
p_i(\phi,\bet,\th)
&=&
\trace\ M_\iout(\phi) \sig(\th,\bet),\
i=1,\ldots,\nout
\\
\sig(\th,\bet) &=& \sumts_k Q_k(\th)\rho(\bet) Q_k(\th)^\dag
\eea
\eeq
The state $\sig(\th,\bet)$ is the output of the quantum channel
$Q(\th)$ and the input to the POVM.  Suppose that the input and POVM
configuration parameters can be selected, respectively, from
$\set{\bet_\ell}{\ell=1,\ldots,\ninput}$ and
$\set{\phi_k}{k=1,\ldots,\npovm}$. 
Hence, under the stated conditions, the worst-case experiment
design problem \refeq{wca} becomes,
\beq[eq:wcs phirho]
\bea{ll}
\mbox{maximize}
&
{\ds\min_{r=1,\ldots,\nth}} 
\sumts_{k=1}^\npovm\sum_{\ell=1}^\ninput\ \lam_{k\ell}\ 
g(\phi_k,\bet_\ell,\th_r)
\\
\mbox{subject to}
&
\lam_{k\ell} \geq 0,\
\sumts_{k=1}^\npovm\sum_{\ell=1}^\ninput\lam_{k\ell} = 1
\\
&
g(\phi,\bet,\th)
=
\sumts_{i=1}^\nout
\left(
\nabla_\th\ p_i(\phi,\bet,\th)
\right)^2
/p_i(\phi,\bet,\th)
\eea
\eeq
Similarly, the average-case experiment design problem \refeq{aca}
becomes,
\beq[eq:acs phirho]
\bea{ll}
\mbox{maximize}
&
\sumts_{k=1}^\npovm\sum_{\ell=1}^\ninput\lam_{k\ell}\ \gav(\phi_k,\bet_\ell)
\\
\mbox{subject to}
&
\lam_{k\ell} \geq 0,\
\sumts_{k=1}^\npovm\sumts_{\ell=1}^\ninput\lam_{k\ell} = 1
\\
&
\gav(\phi_k,\bet_\ell) = \sum_{r=1}^\nth p(\th_r)g(\phi_k,\bet_\ell,\th_r)
\eea
\eeq
The worst-case distribution, $\lam^{\rm wc}$, is obtained by solving
the LP \refeq{wcs phirho}. Following \refeq{lam acs1}, the
average-case distribution, $\lamac$, which solves \refeq{acs phirho}
is explicitly,
\beq[eq:lam acs phirho]
\lam^{\rm ac}_{k\ell}
=
\left\{
\bea{ll}
1 
& 
k,\ell = {\ds\arg\max_{k',\ell'}}\
\gav(\phi_{k'},\bet_{\ell'})
\\
0 & \mbox{otherwise}
\eea
\right.
\eeq
Solutions to \refeq{wcs phirho} and \refeq{acs phirho}, respectively,
$\lamwc$ and $\lamac$, can be used to evaluate the worst-case and
average-case levels of Fisher information as a function of the
uncertain parameter $\th\in\Th$:
\begin{eqnarray}
F(\lam^{\rm wc},\th) 
&=&
\sumts_{k=1}^\npovm\sumts_{\ell=1}^\ninput\ \lam^{\rm wc}_{k\ell}\ 
g(\phi_k,\bet_\ell,\th)
\label{eq:fth wc}
\\
F(\lam^{\rm ac},\th) 
&=&
\lam^{\rm ac}_{k\ell}\ g(\phi_k,\bet_\ell,\th)
\label{eq:fth ac}
\end{eqnarray}
In addition, as benchmarks for \emph{each} $\th\in\Th$, we can compute
the maximum possible, subject to the constraints on the input and
measurement scheme, and the QFI which is the maximum possible with
\emph{no measurement constraints}: the POVMs do not depend upon a
configuration parameter as in \refeq{pout}.  The maximum subject to
the constraints is,
\beq[eq:fth max]
F_{\max}(\th) = \max_{k,\ell} g(\phi_k,\bet_\ell,\th)
\eeq
For the single parameter system of Figure \ref{fig:qsys}, the QFI is
given by, \cite{Holevo:82,BraunsteinC:94},
\beq[eq:fth qfi]
\bea{rcl}
&&
F_{\rm QFI}(\th,\bet) = \trace\ S(\th,\bet)^2\sig(\th,\bet)
\\
&&
S(\th,\bet)\sig(\th,\bet)+\sig(\th,\bet)S(\th,\bet)
=
2\nabla_\th\ \sig(\th,\bet)
\eea
\eeq
with $\sig(\th,\bet)$ from \refeq{pout} and $S(\th,\bet)$ the solution
to the above (matrix) Lyapunov equation. $F_{\rm QFI}(\th,\bet)$,
generally depends on the unknown parameter value $\th$, and in this
case also on the input configuration parameter $\bet$. As developed in
\cite{Holevo:82,BraunsteinC:94,SarovarM:06,GLM:06,BFCG:07,ShajiC:07},
for a unitary channel of the form $U(\th)=\exp(-i\th H_0)$, there is a
$\th$-dependent pure state input $\ket{\psi(\th)}$ such that the QFI
is explicitly,
\beq[eq:fqfi pure]
F_{\rm QFI}(\th) 
= 
\left( \lam_{\max}(H_0)-\lam_{\min}(H_0) \right)^2
\eeq
with $\lam_{\max},\ \lam_{\min}$ here denoting the maximum and minimum
eigenvalues of the Hamiltonian $H_0$. 

We ought to mention that the form of the system shown in Figure
\ref{fig:qsys} is not the most general. For example, the ``OSR'' block
might depend jointly on both $\th$ and a configuration parameter
$\alf$.  The method, however, remains the same.

\section{Example: perturbed unitary channel}
\label{sec:example}

To illustrate the optimization methods we assume the quantum channel
in Figure \ref{fig:qsys} is a unitary channel whose output is
corrupted by \emph{amplitude damping}.  The unitary part is
$U(\th)=\exp(-i\th H_0)$, with
$
H_0 = \frac{1}{\sqrt{2}}\tiny{\left[\bea{cc} 1&1\\1&-1\eea\right]}
$ 
and with the unknown parameter $\th$ uniformly distributed in the
set,
$
\Th=\set{\th}{0.2\leq\th/(\pi/2)\leq 0.8}
$.
The amplitude damping channel can be described by an OSR with two
elements (see, \eg, \cite{NielsenC:00}),
$
A_1(\gam) = \tiny{\left[\bea{cc} 1&0\\ 0& \sqrt{1-\gam}\eea\right]},
$
$
A_2(\gam) = \tiny{\left[\bea{cc} 0&\sqrt{\gam}\\ 0&0\eea\right]}
$
with $\gam$ the probability of dissipation.  It follows that the OSR
of $Q(\th)$ in Figure \ref{fig:qsys} has two elements, $Q_k(\th) =
A_k(\gam)U(\th),\ k=1,2$.

The available input for the experiment is the $2\times 1$ pure state
$\ket{\psi(\bet)}$ which can be adjusted via an angle $\bet$ as:
$
\ket{\psi(\bet)} = \cos\bet\ket{0}+\sin\bet\ket{1},
\;
0 \leq \bet \leq \pi
$.
The POVMs can be adjusted via an angle $\phi$ as:
\[
\left.
\bea{rcl}
M_1(\phi) &=& \ket{z(\phi)}\bra{z(\phi)}
\\
M_2(\phi) &=& I_2 - M_1(\phi)
\\
\ket{z(\phi)} &=& \cos\phi \ket{0}+\sin\phi \ket{1}
\eea
\right\}
\;
0 \leq \phi \leq \pi
\]
We determine the Fisher information for two amplitude damping
probabilities: $\gam\in\seq{0,\ 0.25}$ with $\nth=100$ uniformly
spaced samples of $\th\in\Th$. The POVM and input configuration angles
$\bet,\phi$ are selected from their allowable ranges with $\ninput=10$
and $\npovm=10$ uniformly spaced samples for each of the following
three configuration constraints:
\ben
\item POVM configured $(0\leq\phi\leq\pi)$, input fixed $(\bet=0)$
\item POVM fixed $(\phi=0)$, input configured $(0\leq\bet\leq\pi)$
\item POVM \& input configured $(0\leq\bet\leq\pi,\ 0\leq\phi\leq\pi)$
\een

\begin{figure}[thb]
\hspace{-3ex}
\epsfig{file=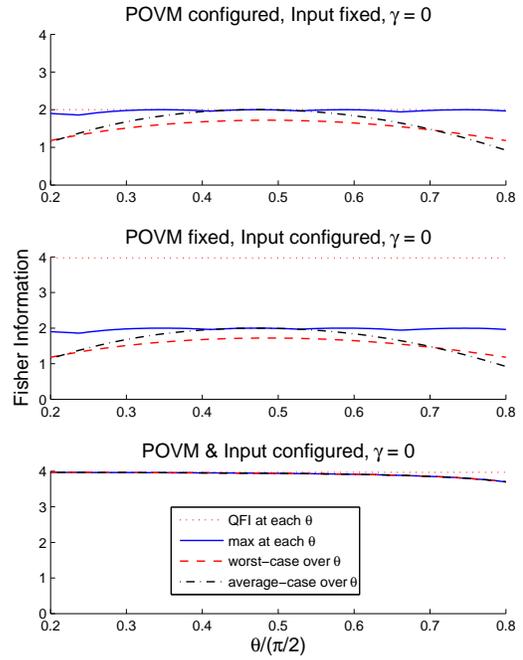,height=3.75in}
\vspace{-5ex}
\caption{Comparison of configuration constraints with $\gam=0$}
\label{fig:comp0p00}
\end{figure}

\begin{figure}[thb]
\hspace{-3ex}
\epsfig{file=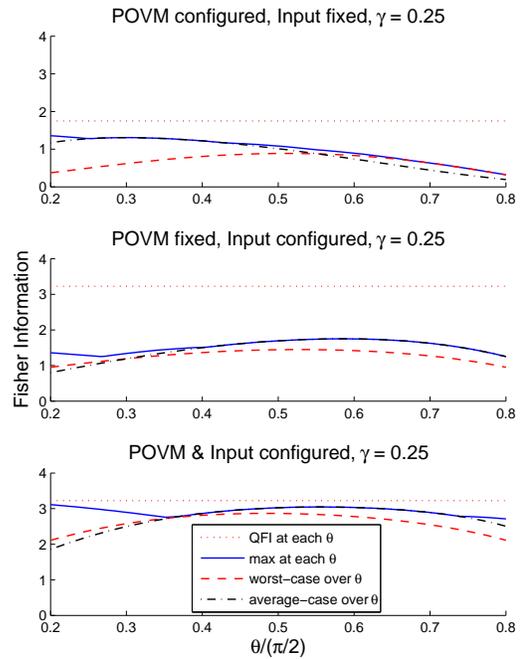,height=3.75in}
\vspace{-5ex}
\caption{Comparison of configuration constraints with $\gam=0.25$}
\label{fig:comp0p25}
\end{figure}

\noindent
Figures \ref{fig:comp0p00}-\ref{fig:comp0p25} show the Fisher
information as a function of the parameter $\th$ for the two values of
amplitude damping and the three configuration constraints.  In each
figure the dotted lines are the QFI for each $\th$ \refeq{fth
qfi}. Note that the absolute maximum for the QFI is achieved only for
$\gam=0$ (unitary channel) and using \refeq{fqfi pure} with $H_0$ as
given above gives $F_{\rm QFI}(\th)=4$.  The solid lines are the
maximum achievable for each value of $\th$ that maximizes the Fisher
information under the configuration constraints \refeq{fth max}. The
dashed lines are what is achieved by using the worst-case distribution
of experiments \refeq{fth wc}, and the dot-dash lines are the
average-case distribution of experiments \refeq{fth ac}.

In all cases, the constrained Fisher information $F(\lam^\ac,\th)$,
and $F(\lam^\wc,\th)$ are relatively close, sometimes nearly
coincident to the maximum possible, $F_{\rm max}(\th)$, and all are
lower than the QFI. When both POVM and input are jointly configured
the constrained information begins to approach $F_{\rm max}(\th)$.
The curves for the case where only the POVM is configured are
generally below those where only the input is configured.

\begin{table}[t]
\centering
\btab{|c|c||ccc|ccc|}
\hline
$\gam$ & Configured & \multicolumn{3}{c|}{Average-Case} 
& \multicolumn{3}{c|}{Worst-Case}\\
&& $\phi/\pi$ & $\bet/\pi$ & $\lamac$ 
& $\phi/\pi$ & $\bet/\pi$ & $\lamwc$ \\
\hline\hline
0 & POVM & .89 & 0 & 1 & .44 & 0 & .57\\
\cline{2-8}
& Input & 0 & .89  & 1 & 0 & .44 & .57\\
&&&&& 0 & .78 & .43\\
\cline{2-8}
& POVM \& Input & .89 & .89 & 1 & .89 & .89 & .89\\
\hline
0.25 & POVM & .44 & 0  & 1 & .78 & 0 & 1\\
\cline{2-8}
& Input & 0 & .33 & 1 & 0 & 0 & .14\\
&&&&& 0 & .33 & .72\\
&&&&& 0 & 1 & .14\\
\cline{2-8}
& POVM \& Input & .89 & .33 & 1 & .89 & .33 & .80\\
&&&&& .89 & .89 & .20\\
\hline
\etab
\vspace{1ex}
\caption{Optimal distributions}
\label{tab:lam}
\end{table}
The numerically non-zero elements of the worst-case and average-case
optimal distributions for all the cases are shown in Table
\ref{tab:lam}. By construction, only one input configuration is
required for the average-case distribution \refeq{lam acs phirho}. The
worst-case distribution requires up to 3 configurations when
$\gam=0.25$. In this example the configuration angles remain
relatively unchanged exhibiting some robustness to the amplitude
damping probability $\gam$. The worst-case distributions change more
significantly. Given the relatively close levels of Fisher
information for $\th\in\Th$, it would seem more prudent to use the
single-setting of input and POVM obtained from the average-case
optimization. In most experiments there is a penalty in terms of time
to reset the configurations. 

\begin{figure}[t]
\centering
\epsfig{file=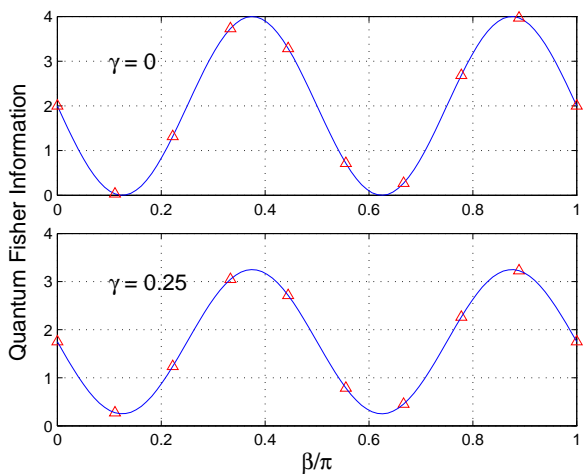,width=3.5in}
\vspace{-4ex}
\caption{QFI vs. input configuration parameter
$\bet$ for $\gam\in\seq{0,0.25}$.  
\red{$\triangle$} are the $\ninput=10$ available.}
\label{fig:qfish}
\end{figure}

In this example the available input configurations effect the QFI. The
solid lines in Figure \ref{fig:qfish} show the QFI for each value of
$\gam$ vs. the input configuration angle $\bet$ for a large number of
samples in the range. The QFI in this case is independent of $\th$.
The triangles show the $\ninput=10$ available values.  The solid lines
indicate that multiple inputs can achieve the bound whereas the
restricted set forces a unique maximum which does not necessarily
occur at the true maximum. For example, as seen in the top plot for
$\gam=0$, the constrained maximum is near the global maximum ($F_{\rm
QFI}(\th)=4$). This is achieved only in the case with $\gam=0$ and
clearly over bounds the plot for $\gam=0.25$. As might be expected, a
perturbation of the unitary channel, in this case via amplitude
damping, makes it harder to attain the maximum possible QFI.  Observe
also that if the inputs were further constrained, say
$\bet/\pi\in\seq{0,0.2,0.5,0.8}$, then the achieved QFI would not be
nearly as close to the maximum possible. The analysis of this examples
thus provides the designer with information about the limit of
performance of the system. If the potential performance increase over
what is available under the constraints on instrumentation is
significant, then a more flexible instrumentation might be considered
worthwhile.

\vspace{-3ex}
\section{Conclusion}
\label{sec:conclude}

We have shown that maximizing the precision in estimating a single
parameter in a quantum system subject to input and POVM constraints
reduces to a linear program for both what is defined here as a
worst-case and average-case objective. For the average-case, the
solution to the linear program can be expressed analytically and
involves a simple search, \ie, find the largest element of an easily
computed vector. Both solutions provide different levels of Fisher
information over the range of anticipated parameter
variation. Comparing these constrained solutions to the best possible
under the constraints as well as to the QFI gives an indication of the
performance limitations imposed by the constraints.

Future efforts will consider the effect of entanglement and
multi-parameter estimation. 

\bibliographystyle{prsty}
\bibliography{D:/robert/tex/rlk}

\begin{thebibliography}{10}

\bibitem{Holevo:82}
A.~S. Holevo, {\em Probabilistic and Statistical Aspects of Quantum Theory}
  (North-Holland, Amsterdam, 1982).

\bibitem{BraunsteinC:94}
S. Braunstein and C. Caves, Phys. Rev. Lett. {\bf 72},  3439  (1994).

\bibitem{SarovarM:06}
M. Sarovar and G.~J. Milburn, J. Phys. A: Math. Gen. {\bf 39},  8487  (2006).

\bibitem{GLM:06}
V. Giovannetti, S. Lloyd, and L. Maccone, Phys. Rev. Lett. {\bf 96},  010401
  (2006).

\bibitem{BFCG:07}
S. Boixo, S. Flammia, C. Caves, and J. Geremia, Phys. Rev. Lett. {\bf 98},
  090401  (2007).

\bibitem{ShajiC:07}
A. Shaji and C.~M. Caves, Phys. Rev. A {\bf 76},  032111  (2007).

\bibitem{Cramer:46}
H. Cram\'{e}r, {\em Mathematical Methods of Statistics} (Princeton Press,
  Princeton, NJ, 1946).

\bibitem{BoydV:04}
S. Boyd and L. Vandenberghe, {\em Convex Optimization} (Cambridge University
  Press, Cambridge, UK, 2004).

\bibitem{KosutWR:04}
R.~L. Kosut, I.~A. Walmsley, and H. Rabitz, quant-ph/0411093  (2004).

\bibitem{NielsenC:00}
{M.A. Nielsen and I.L. Chuang}, {\em {Quantum Computation and Quantum
  Information}} ({Cambridge University Press}, Cambridge, UK, 2000).

\end{thebibliography}

\end{document}